\documentclass[%
 reprint,
 amsmath,amssymb,aps,twocolumn,
pdflatex
]{revtex4}
\usepackage{epsfig}
\usepackage{graphicx}
\usepackage{dcolumn}
\usepackage{bm}
\usepackage[utf8]{inputenc}
\usepackage[T1]{fontenc}
\usepackage{mathptmx}
\usepackage{xcolor}
\begin{document}
\preprint{AIP/123-QED}
\title[]{
T-wave Inversion through Inhomogeneous Voltage Diffusion within the FK3V Cardiac Model 
}

\author{E. Angelaki$^{1, 2}$, N. Lazarides$^{3}$, G. D. Barmparis$^{1}$, 
        Ioannis Kourakis$^{3}$, Maria E. Marketou$^{4,5}$, G. P. Tsironis$^{1, 2}$
}
\affiliation{$^{1}$Department of Physics, and 
                   Institute of Theoretical and Computational Physics, 
                   University of Crete, Heraklion, Greece} 
\affiliation{$^{2}$Harvard John A. Paulson School of Engineering and Applied Sciences, 
                   Harvard University, Cambridge, MA, USA 
}
\affiliation{$^{3}$Department of Mathematics, Khalifa University of Science and Technology, 
                   P.O. Box 127788, Abu Dhabi, United Arab Emirates} 
\affiliation{$^{4}$School of Medicine, University of Crete, Heraklion,  Greece}
\affiliation{$^{5}$Department of Cardiology, Heraklion University Hospital, Heraklion, Greece}
%%%%%%%%%%%%%%%%%%%%%%%%%%%%%%%%%%%%%%%%%%%%%%%%%%%%%%%%%%%%%%%%%%%%%%%%%%%%%%%%
%%\date{\today}
%%%%%%%%%%%%%%%%%%%%%%%%%%%%%%%%%%%%%%%%%%%%%%%%%%%%%%%%%%%%%%%%%%%%%%%%%%%%%%%%
\begin{abstract}
\noindent The heart beats due to the synchronized contraction of cardiomyocytes 
triggered by a periodic sequence of electrical signals called action potentials, 
which originate in the sinoatrial node and spread through the heart's electrical 
system. A large body of work is devoted to modeling the propagation of the 
action potential and to reproducing reliably its shape and duration. Connection 
of computational modeling of cells to macroscopic phenomenological curves such 
as the electrocardiogram has been also intense, due to its clinical importance 
in analysing cardiovascular diseases. In this work we simulate the dynamics of 
action potential propagation using the three-variable Fenton-Karma model that
can account for both normal and damaged cells through spatially inhomogeneous 
voltage diffusion coefficient.  We monitor the action potential propagation in 
the cardiac tissue and calculate the pseudo-electrocardiogram that reproduces the 
R and T waves. The R wave amplitude varies according to a double exponential law 
as a function of the (spatially homogeneous, for an isotropic tissue) diffusion 
coefficient. The addition of spatial inhomogeneity in the diffusion coefficient by 
means of a defected region representing damaged cardiac cells, may result in T-wave 
inversion in the calculated pseudo-electrocardiogram. The transition from positive 
to negative polarity of the T-wave is analyzed as a function of the length and the 
depth of the defected region.
\end{abstract}
\date{\today}
\maketitle

%%%%%%%%%%%%%%%%%%%%%%%%%%%%%%%%%%%%%%%%%%%%%%%%%%%%%%%%%%%%%%%%%%%%%%%%%%%%%%%%
% Lead Paragraph.
\noindent {\bf Cardiovascular diseases (CVDs) are the leading cause of death 
globally. The healthy heart produces a synchronized mechanical contraction by a 
self-generated electrical signal that propagates through the muscle as an action 
potential (AP) wave, and is tracked through the electrocardiogram (ECG), perhaps 
the most widely used clinical tool for the detection and diagnosis of a broad 
range of cardiac conditions. The last decades, there has been a fast growth of 
sophisticated and detailed mathematical models that encompass realistic 
electrophysiological and anatomical properties, aiming to help understanding 
life-threatening situations related to CVDs and developing appropriate therapies. 
Moreover, significant progress has been achieved in solving the forward problem 
of electrocardiography to obtain the simulated ECG (i.e., the pseudo-ECG) of a 
patient. The detailed computational models, however, often defy simple 
mathematical analysis and transparency; fortunately, the more flexible 
lower-dimensional phenomenological (``simple'') models, such as the three-variable 
Fenton-Karma (FK3V) model, can reproduce quantitatively the overall characteristics 
of cardiac tissue that are relevant to the AP propagation. Importantly, a 
pseudo-ECG can be calculated from the action potentials obtained from the FK3V 
model, which reproduce some features of observed ECGs. Specifically, they reproduce 
the R and T wave, as well as the T-wave inversion commonly found in myocardial 
ischemia. In that condition, the electric conductance in a particular region of 
the cardiac tissue (scar) is severely reduced. Within the FK3V model, this 
situation can be described by inhomogeneous voltage diffusion coefficient with 
very low value in the region of the scar. For sufficiently large scar, the 
polarization of the T wave is inverted from positive to negative, providing thus 
a strong link between a phenomenological quantity and measured data. That link 
of the inverted T-wave and the inhomogeneous diffusion coefficient (i.e., the 
diffusion coefficient with a defected region representing a scar) may be used 
in identifying the location and the width of the scar by solving the inverse 
problem, which may be a matter of future research. 
}

%%%%%%%%%%%%%%%%%%%%%%%%%%%%%%%%%%%%%%%%%%%%%%%%%%%%%%%%%%%%%%%%%%%%%%%%%%%%%%%%
\section{Introduction}
\noindent The heart is a muscular organ situated between the right and left lungs 
whose primary role is to pump oxygen-rich blood throughout the body. It has four 
main chambers; the two smaller upper ones are called atria, and the larger lower 
ones are called ventricles. Life is sustained due to the reliable propagation 
of action potentials (AP) across the cardiac muscle, or myocardium, which 
ensures its coordinated excitation and contraction, i.e., the heartbeat. 
The AP is essentially an electrical disturbance, which propagates over long 
distances preserving its amplitude. Once initiated by excitation from a stimulus
current, its propagation becomes independent of the triggering stimulus, 
achieving thus an ``autopreserving'' status. To initiate the AP, the triggering 
stimulus current must assume a threshold value of certain amplitude and duration. 

%%%%%%%%%%%%%%%%%%%%%%%%%%%%%%%%%%%%%%%%%%%%%%%%%%%%%%%%%%%%%%%%%%%%%%%%%%%%%%%%
\begin{figure*}[!t]
   \includegraphics[scale=0.45]{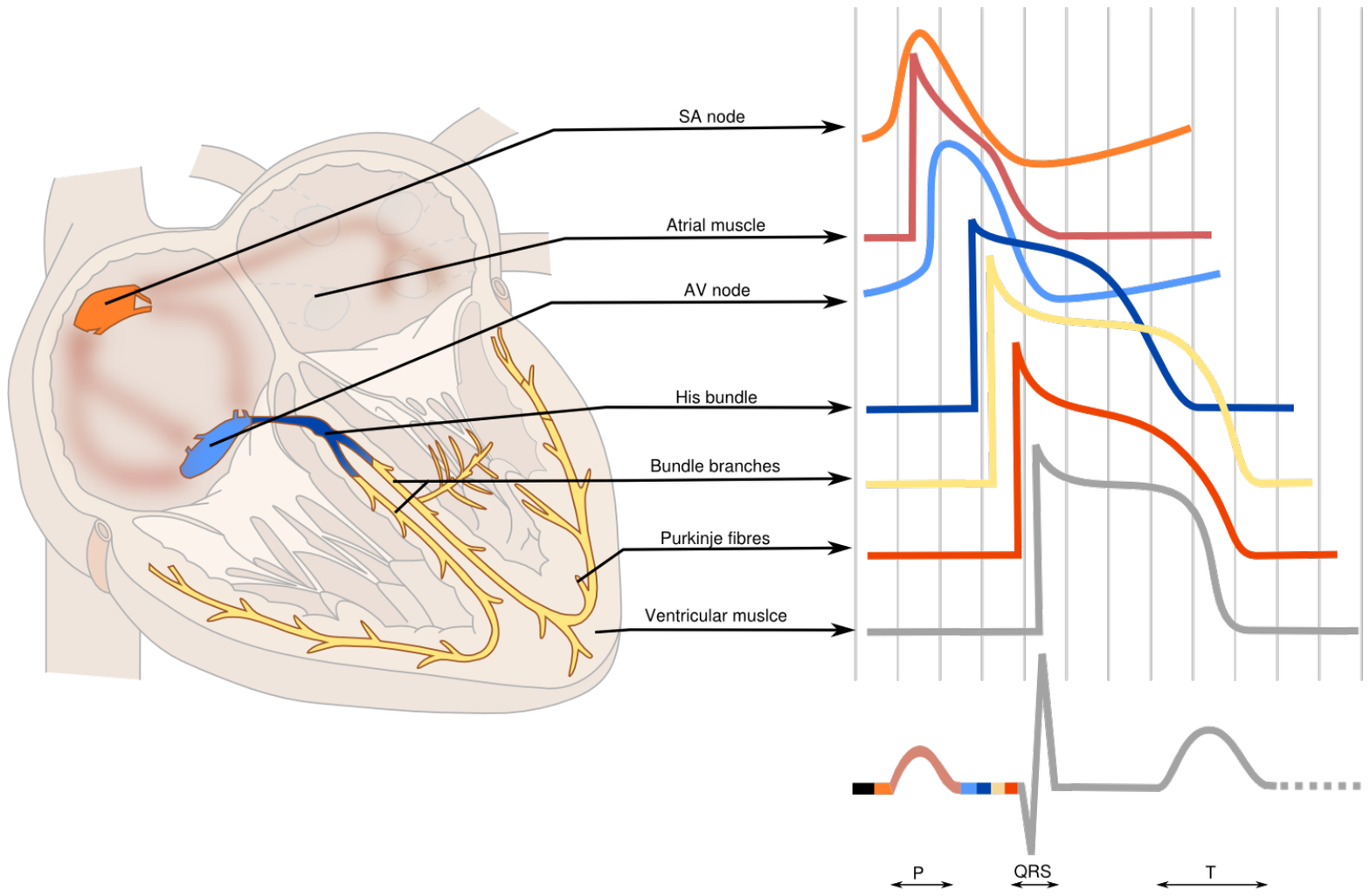}
   \caption{Cardiac action potential shown for different types of cardiomyocytes, 
            whose names are shown on the arrows, and how they relate to the 
            electrocardiogram.  
            Image courtesy of Dr. De Voogt and ECGpedia.org
}
\label{fig:Conduction}
\end{figure*}
%%%%%%%%%%%%%%%%%%%%%%%%%%%%%%%%%%%%%%%%%%%%%%%%%%%%%%%%%%%%%%%%%%%%%%%%%%%%%%%%

\indent Cardiac cells, called myocytes, are roughly shaped as cylinders 
$100 \mu$m long and $25 \mu$m wide \cite{CE}, and are metabolically and 
electrically connected via gap junctions \cite{Kleber2021}. Action potential 
propagation involves the diffusion of ions between cells via these gap junctions, 
as well as their transfer in and out of the cell via transmembrane ionic currents. 
Effectively, gap junctions slow down propagation by having a larger resistance 
than the cytoplasm. Gap junctions allow heart cells to function in a coordinated, 
synchronized manner, ensuring they are electrically connected as a single unit. 
These junctions are predominantly found at the ends of cells. As a result, the 
anatomic characteristics of groups of cardiac muscle differ based on the 
orientation they are studied from, a trait known as ``anisotropy''. Conduction 
velocity is typically faster, about two to three times, along the length of the 
fiber compared to across its width.

Cardiac electrical propagation is often modeled as a reaction-diffusion 
process. Ionic models that describe this process have become increasingly more 
complex and more realistic. For the ventricular AP across mammalian cardiac 
cells, several ionic models with simplified ionic currents have been developed, 
reviewed in a recent article \cite{Alonso2016}. The mathematical aspects of 
cardiac electrophysiology have been exposed in Ref. \cite{Franzone2014}, while 
the computing aspects in Refs. \cite{Ji2016,Golemati2019}. Two such models are 
the three-variable cardiac Fenton-Karma (FK3V) model \cite{Fenton1998}, and the 
four-variable Bueno-Cherry-Fenton model \cite{Bueno2008},  which have been shown 
previously to be highly useful for reproducing a broad range of dynamics of cardiac 
cells and tissue obtained experimentally or from other more complex models. 
Recently, the FK3V model was used in the reconstruction of cardiac electrical 
excitations from optical mapping recordings \cite{Marcotte2023}, while a
three-dimensional anisotropic version of it was used to simulate vortices in the 
lower heart chambers \cite{Zhang2017}. Moreover, an extension of the FK3V model 
that accounts for stochastic effects has been reported \cite{Marcotte2021}. 
We should also mention the more involved 
electrophysiological models such as the Beeler-Reuter model \cite{Beeler1977}, 
the Luo-Rudy model \cite{Luo1991}, and the TenTusscher-Noble-Noble-Panfilov model 
\cite{TenTusscher2004}, which are based on direct experimental observations. 
These models, though, are too complex to provide an essential phenomenological 
insight into the spatial dynamical behavior of the AP, and this is the reason 
we chose the FK3V model.

A macroscopic depiction of the cardiac electrical activity is the electrocardiogram
(ECG) tracing, a relatively inexpensive and widely available medical screening tool. 
Recorded using a machine called an electrocardiograph, it is the primary evaluation 
method for a person suspected of having a cardiovascular pathology 
\cite{SiontisKonstantinosC2021Aiei,AngelakiEleni2021Doal}. Analysis of the ECG using 
various methods has helped clinicians detect cardiovascular abnormalities 
\cite{HaganRachael2021Coml}, e.g., one study uses a single (out of the usually 
acquired $12$) lead in an ECG, to detect T-wave (ventricular repolarisation) morphology 
abnormalities \cite{TutukoBambang2022DEls}.

Fig.\;\ref{fig:Conduction} shows the cardiac conduction system as a network of 
specialized cells comprising of the sinoatrial node, the atrial muscle, the 
atrioventricular node, the His bundle and its bundle branches, the Purkinjie 
fibers, and finally the ventricular myocytes. Depicted is the membrane potential 
with respect to time, for the duration of a single heartbeat. Note that AP curve 
shapes are different for each type of cardiac cell. P waves relate to the 
depolarization of the atrial myocytes, the QRS complex relates to the 
depolarization of the ventricles, and T waves relate to the repolarization of 
the ventricles. We notice that, the AP of the ventricular cells, depicted by the 
grey curve at the bottom of the series of curves, has a longer duration than that 
of the sinoatrial node, drawn as the top curve; also, the Purkinje cell AP is 
similar to the ventricular action potential except for a sharper initial peak. 
These relations are color-coded in the small realistic ECG at the bottom right 
of the image. Disruptions in AP propagation are the manifestations of underlying 
cardiac abnormalities; in myocardial ischemia, for example, the blood supply to 
the heart's coronary arteries cannot meet the demand.

%%%%%%%%%%%%%%%%%%%%%%%%%%%%%%%%%%%%%%%%%%%%%%%%%%%%%%%%%%%%%%%%%%%%%%%%%%%%%%%%
The basis of ischemic arrhythmogenesis is the alteration in the electrical 
properties of ventricular tissue, producing changes in the AP pulse morphology 
and the body surface ECG \cite{WitAndrewL1993Tvao,2018Ce:f}. One such alteration, 
the remodeling of ionic currents due to changes in intracellular and 
extracellular ionic concentrations, has been studied in the literature 
\cite{ShawRM1997Eeoa}. In addition to ionic remodeling, spatial heterogeneity 
such as cell-to-cell decoupling, occurring usually in later stages of ischemia, 
has been shown experimentally to lead to propagation disruptions and a reduction 
in conduction velocity \cite{JongsmaH.J2000Gjic}.

In this work, we perform extensive simulations using the FK3V model for the 
ventricular AP over a one-dimensional (1D) cable transversal to the ventricular 
tissue. The obtained AP is then used to generate a related ECG pattern, usually 
called a pseudo-ECG \cite{Aslanidi2005,Wang2006}, whose morphology under 
different values and profiles of the voltage diffusion coefficient is then 
explored. We mimic a spatially localized area of depressed conductivity by 
reducing the diffusion coefficient considerably in that specific area. Within the 
framework of the FK3V model we were able to reproduce the R and T waves of the 
ECG through appropriately adjusting the characteristics of the stimulus current. 
Moreover, the calculated pseudo-ECG exhibits T-wave inversion which may become 
deep for relatively large scar tissue areas in the heart, in a way resembling 
what has been observed for patients with myocardial ischemia. For simplicity, 
only one set of electrophysiological parameters is used, i.e., the cable spans 
a single region of the ventricular tissue.

One-dimensional numerical simulations, being quick and efficient, enabled us 
to try out multiple different values for the relevant parameters and capture the 
changes in morphology. The aforementioned property of conduction velocity being 
typically about two to three times faster along the length of the fiber compared 
to across its width, makes numerical calculations using 1D models a good first 
approach.

%%%%%%%%%%%%%%%%%%%%%%%%%%%%%%%%%%%%%%%%%%%%%%%%%%%%%%%%%%%%%%%%%%%%%%%%%%%%%%%%
\section{Methods}
\subsection{The three variable model by Fenton and Karma}
Computational models allow the study of AP propagation in single cells, in 1D 
cables of cells, in two-dimensional slabs of tissue, as well as in 
three-dimensional whole heart models. The FK3V model of coupled reaction-diffusion 
equations on a 1D cable of cells is used in this work to produce pseudo-ECG 
patterns relating to AP propagation. We present the equations briefly here; for 
a detailed presentation one may consult the original article by Fenton and Karma 
\cite{Fenton1998} or the review article by Alonso {\em et al.} \cite{Alonso2016}.

Our theoretical cable of cells, of length $L=3\;$cm, is composed of $400$ 
ventricular cells of a single cell type, connected via gap junctions. A stimulus 
current $J_{\text{stim}}(x,t)$ with an above-threshold amplitude is applied to the
first $15$ cells (i.e., with the first cell being at $x=0$). That current is 
therefore assumed to excite a small, spatially restricted region around the left 
end of the cable of length $L_{\text{exc}}=15 \times \text{<cell length>}$. 
We take the cardiac cell length to be equal to the spatial discretization 
$dx=0.0075$ cm, so that $L_{\text{exc}} \simeq 0.11$ cm. In what follows, 
the stimulus current $J_{\text{stim}}$ is taken to be a rectangular pulse
of amplitude $J_{\text{amp}}=0.9$ mA and duration $\tau_{\text{p}} = 11$ ms, 
unless otherwise stated. The model consists of the three coupled partial 
differential equations 
(for completeness the stimulus current is also included)
\begin{align}
 \label{eq:eq01}
   \frac{\partial u}{\partial t} &=\nabla \cdot \left( \tilde{D} \nabla u \right)
 -J_{\text{fi}}(u;v) -J_{\text{so}}(u) -J_{\text{si}}(u;w) + J_{\text{stim}}(x, t) \\
\label{eq:eq02}
   \frac{\partial v}{\partial t} &=\Theta(u_c-u) \frac{1-v}{\tau_{\text{v}}^-(u)}
   -\Theta (u-u_{\text{c}}) \frac{v}{\tau_{\text{v}}^+}\\
\label{eq:eq03}
   \frac{\partial w}{\partial t} &=\Theta(u_{\text{c}}-u) \frac{1-w}{\tau_{\text{w}}^-}
-\Theta(u-u_c) \frac{w}{\tau_{\text{w}}^+}
\end{align}

The normalized transmembrane voltage function $u(x,t)$ is obtained through the relation
\begin{equation}
\label{eq04_0}
   u(x,t) \equiv \frac{V(x,t) -V_0}{V_{fi} -V_0},
\end{equation}
where $V(x,t)$ is the un-normalized transmembrane potential measured in units of mV, 
$V_0$ is the resting membrane potential, and $V_{fi}$ is the Nernst potential of the 
fast inward current. The normalized threshold potential is given by $u_c$.

The permeability of the channels in the cell membrane is regulated by the two gating 
variables $\nu(x,t)$ and $w(x,t)$. Gate state indicates whether ions can pass through 
the membrane or not. The variable $\nu(x,t)$ denotes the fast inactivation gate which 
opens when the cell is not excited, and closes when it becomes excited. The closing 
time constant $\tau_v^+$ corresponds to cell depolarization, and the opening time 
constant $\tau_v^-$ to cell repolarization. The $u-$dependent parameter $\tau_v^-(u)$ 
is given by
\begin{equation}
\label{eq04}
   \tau_{\text{v}}^-(u) =\Theta(u -u_{\text{v}}) \tau_{\text{v1}}^- 
                         +\Theta(u_{\text{v}} -u) \tau_{\text{v2}}^-. 
\end{equation}
This splitting allows the minimum diastolic interval, i.e., the excitable gap, 
controlled by $\tau_{v1}^-$, to vary independently from the steepness of this 
curve, controlled by $\tau_{v2}^-$. The voltage threshold $u_c > u_{\nu}$ 
controls the splitting. The variable $w$ is the probability of a gate opening as 
described in the Hodgkin-Huxley model \cite{Hodgkin1952}; $\tau_w^+$ and 
$\tau_w^-$ are the time constants for closing and opening of the gate, 
respectively.
%%%%%%%%%%%%%%%%%%%%%%%%%%%%%%%%%%%%%%%%%%%%%%%%%%%%%%%%%%%%%%%%%%%%%%%%%%%%%%%%
\begin{table}[!t] 
\centering
\begin{tabular}{||l|c|c|c||} 
\hline \hline
 Parameter      & BR model & MBR model & MLR-I model \\
\hline
 $\bar{g}_{\text{fi}}$ & 4     & 4     &5.8 \\                                               
 $\tau_{\text{r}}$       & 33.33 & 50    &130 \\                                               
 $\tau_{\text{si}}$    & 29    & 44.84 & 127\\                                               
 $\tau_0$       & 12.5  & 8.3   & 12.5 \\                                               
 $\tau_{\text{v}}^+$     & 3.33  & 3.33  &10 \\                                               
 $\tau_{\text{v1}}^-$  & 1250  & 1000  &18.2 \\                                               
 $\tau_{\text{v2}}^-$  & 19.6  & 19.2  &18.2 \\                                               
 $\tau_{\text{w}}^+$     & 870   & 667   & 1020 \\                                               
 $\tau_{\text{w}}^-$     & 41    & 11    & 80\\                                               
 $u_{\text{c}}$          & 0.13  & 0.13  &0.13 \\                                               
 $u_{\text{v}}$          & 0.04  & 0.055 & -- \\                                               
 $u_{\text{c}}^{\text{si}}$     & 0.85  & 0.85  &0.85 \\    
\hline                                           
 Other Parameters &  &  &  \\
\hline                                           
 $C_{\text{m}}$          & $1~{\mu}$F/cm$^2$  & & \\                                               
 $V_0$          & $-85~mV$        & & \\                                               
 $V_{\text{fi}}$       & $+15~mV$        & & \\                                                                       
 $k$            & $10$            & & \\                                            
\hline \hline
\end{tabular}
   \caption{Three different sets of model parameters that can be used into the 
            three-variable Fenton-Karma model (Eqs.\;(\ref{eq:eq01})\;-\;(\ref{eq:eq03})). 
            In this work, the parameters of the modified Beeler-Reuter (MBR) 
            model papameters are used (from Ref.\;\cite{Fenton1998E}).
}
\label{table:1}
\end{table}
%%%%%%%%%%%%%%%%%%%%%%%%%%%%%%%%%%%%%%%%%%%%%%%%%%%%%%%%%%%%%%%%%%%%%%%%%%%%%%%%

The scaled phenomenological ionic currents $J_{\text{fi}}$, $J_{\text{so}}$, and 
$J_{\text{si}}$, where the subscript $\text{f}$ means fast and $\text{s}$ slow, 
are related to the corresponding currents in units of mA through 
\begin{equation}
\label{eq05}
   J_{\text{i}} =\frac{I_{\text{i}}}{C_{\text{m}}\;(V_{\text{i}} -V_0)}
\end{equation}
where $C_{\text{m}}$ is the membrane capacitance, and $\text{i}$ represents any 
of the different $\text{fi}$, so, or si. The following remarks on the currents
are worth to be made:

(a) $J_{\text{fi}}$ corresponds to the fast inward sodium (Na$^+$) current, 
responsible for the depolarization of the membrane, and depending on the gating 
variable $\nu$. This gating variable is responsible for inactivation of the 
current after depolarization, and its reactivation after repolarization, 

(b) $J_{\text{so}}$ is a slow outward current analogous to the time-independent 
potassium (K$^+$) current; it is responsible for re-polarization of the cell membrane, 
and 

(c) $J_{\text{si}}$ is a slow inward current, corresponding to the calcium (Ca$^+$) 
current, that balances $I_{so}$ during the plateau phase of the AP; this current 
depends on one gate variable $w$, responsible for its inactivation and reactivation.

The above correspondence to the Na, K, and Ca currents is certainly an 
oversimplification, due to membrane dynamics being a lot more complex. The model, 
though, succeeds in capturing the minimal ionic complexity that underlies the 
membrane recovery processes. All currents are considered normalized. 
The expressions for the normalized currents read
\begin{eqnarray}
\label{eq:eq06}
   J_{\text{fi}}(u;v) =-\frac{v}{\tau_{\text{d}}} \Theta(u-u_{\text{c}}) (1 -u) (u-u_{\text{c}}), 
\\
\label{eq:eq07}
   J_{so}(u)=+\frac{u}{\tau_0} \Theta(u_{\text{c}}-u)  +\frac{1}{\tau_{text{r}}} \Theta(u-u_{\text{c}})
\\
\label{eq:eq08}
   J_{si}(u;w)=-\frac{w}{2 \tau_{\text{si}}} 
      \left\{ 1+\tanh \left[ k \left( u -u_{\text{c}}^{\text{si}} \right) \right] \right\},
\end{eqnarray}
where 
\begin{equation}
\label{eq10b}
   \tau_{\text{d}} =\frac{C_{\text{m}}}{\bar{g}_{\text{fi}}}.
\end{equation}
The values of the parameters $\bar{g}_{\text{fi}}$, $\tau_0$, $\tau_{\text{r}}$, 
$\tau_{\text{si}}$, $k$, and $u_{\text{c}}^{\text{si}}$ are given in Table 
\ref{table:1}. In this work, the values of the modified Beeler-Reuter (MBR) 
model parameters are used in the FK3V equations. The function $\Theta =\Theta(x)$, 
which appears repeatedly in Eqs. (\ref{eq:eq01})-(\ref{eq:eq03}) and 
Eqs. (\ref{eq:eq06})-(\ref{eq:eq08}), is the standard Heaviside step function 
defined by $\Theta(x) =1$ for $x \geq 0$ and $\Theta(x) =0$ for $x < 0$. 
Note that the parenthesis next to the symbol $\Theta$, i.e., 
$\Theta(u -u_{\text{v}})$, is not a multiplicand but the argument of the function.

From Eq.\;(\ref{eq:eq01}) we can see that modeling the propagation of electrical 
impulses in cardiac tissue is affected by two distinct terms. The first term of 
the right hand side, includes the diffusion coefficient and encompasses the 
passive characteristics of the medium, such as its microscopic structure and 
cell-to-cell coupling via ion conducting gap junctions 
\cite{10.1016/j.cardiores.2003.11.035}. 
The second term, the sum of the ionic currents through the membrane channels 
(excluding the stimulus current $J_{\text{stim}}$  ), denotes the dynamic 
characteristic of the medium. As mentioned above, most of the research on 
propagation disruptions concentrates on the remodeling of ionic currents. 
We chose to concentrate on varying the profiles of the voltage diffusion 
coefficient and consequently studying their effect on the calculated pseudo-ECG, 
regarding thus the suppression of electrical connection between cells as the 
primary cause of cardiac pathology. More detailed ionic models may include more 
membrane currents measured in classic voltage-clamp or patch-clamp experiments, 
and a larger number of gates.

%%%%%%%%%%%%%%%%%%%%%%%%%%%%%%%%%%%%%%%%%%%%%%%%%%%%%%%%%%%%%%%%%%%%%%%%%%%%%%%%
\begin{figure}[!h]
    \includegraphics[width=0.4\textwidth]{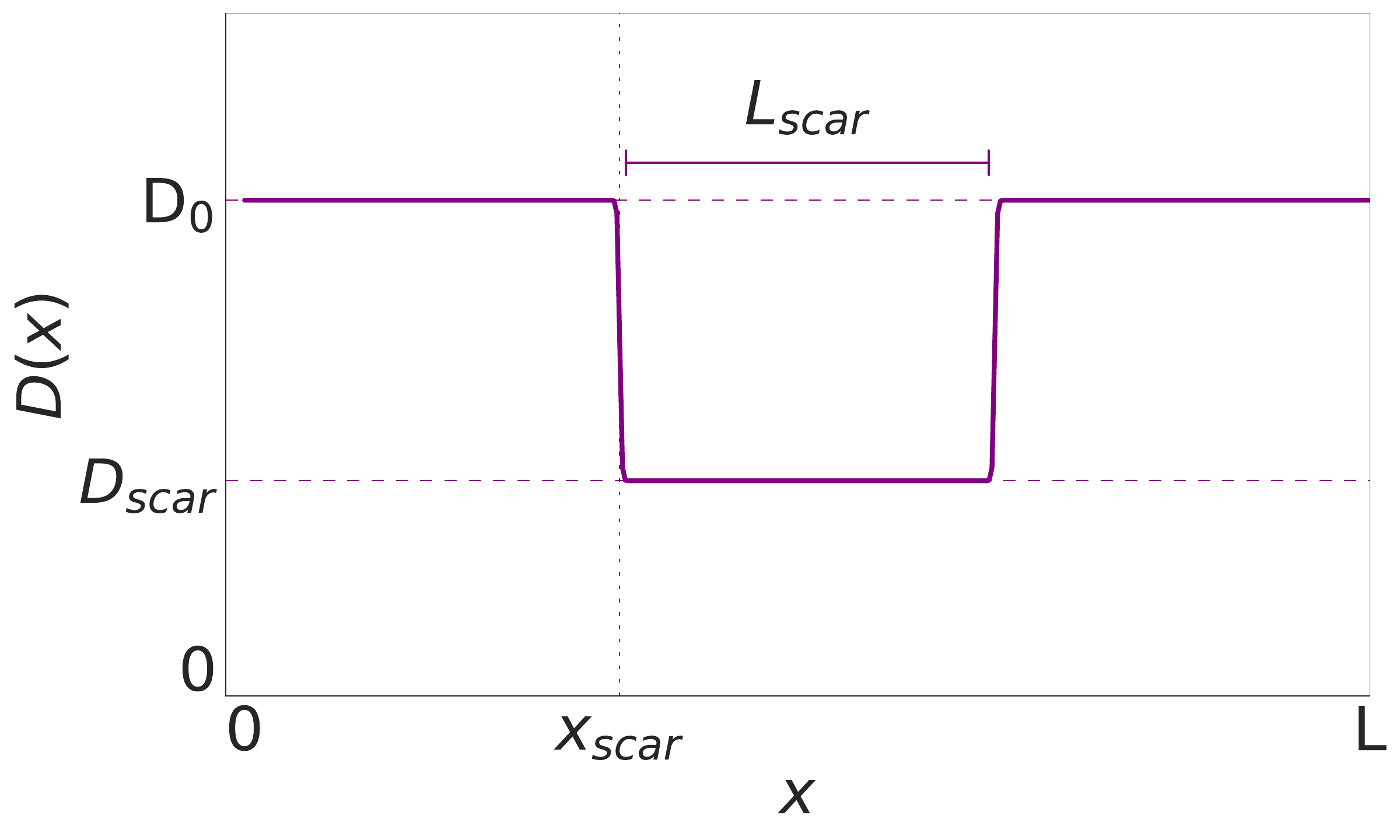}
    \caption{Profile of the diffusion coefficient $D(x) > 0$ ($0 < x < L$) for a 
    localized defect of width $L_{\text{scar}}$ and decrease percentage $\lambda$, 
    around the effective $D_0$ in the regions where the tissue is normal (healthy). 
    The dotted vertical line highlights the start $x_{\text{scar}}$ of the defected
    region (scar). 
    The parameter values used to make this particular plot are 
    $D_0 =0.005$ cm$^2$ms$^{-1}$, $L_{\text{scar}} =1.0$ cm, $x_{\text{scar}} =1.0$ cm, 
    $\lambda=-0.6$, and $L = 3.0$ cm. 
    Using Eq. (\ref{eq:Dscar}) we find $D_{\text{scar}}$ = 0.002 mV.
}
\label{fig:Dtilde}
\end{figure} 
%%%%%%%%%%%%%%%%%%%%%%%%%%%%%%%%%%%%%%%%%%%%%%%%%%%%%%%%%%%%%%%%%%%%%%%%%%%%%%%%

%%%%%%%%%%%%%%%%%%%%%%%%%%%%%%%%%%%%%%%%%%%%%%%%%%%%%%%%%%%%%%%%%%%%%%%%%%%%%%%%
\subsection{The Role of the Diffusion Coefficient}
From the cable equation analysis, the effective voltage diffusion coefficient 
for homogeneous (healthy) tissue is given by
\begin{equation}
\label{eq:D}
   D_0 = \frac{1}{C_m \rho S_u} 
\end{equation}
where $C_m$ is the cell membrane capacitance, $\rho$ is the longitudinal resistivity 
(attributed to the gap junctions), and S$_u$ is the surface-to-volume ratio for 
the cell. The values of the parameters $C_m$\;=\;1\;$\mu$Fcm$^{-2}$, 
$\rho$\;=\;0.4\;k$\Omega$cm (experimentally measured\;\cite{Bueno2008}), and 
$S_u$\;=\;5000\; cm$^{-1}$ provided in Table \ref{table:1} for human ventricular 
cells, give the typical value $D_0\;=0.0012 \pm 0.0002\;$cm$^2$ms$^{-1}$ used often
in literature. In this work we use various values for the diffusion coefficient, 
and by doing so, we can model various sub-cellular characteristics of cardiac 
electrical propagation, such as different gap junction resistance and cell 
membrane capacitance, and by extension, study conduction problems in the heart. 
Myocardial tissue is, of course, a very complex structure, and we hope to 
capture only a part of its behavior.

We study the role of the diffusion coefficient both when it is constant, and when 
it is allowed to vary spatially. In Fig.\;\ref{fig:Dtilde}, we plot the profile 
of the diffusion coefficient $\tilde{D}(x)$ that contains a localized heterogeneity 
in the form of a defected (scar) region in which the conductance velocity has been 
significantly reduced due to reduced electrical connection between cells, i.e., a 
region in which the value of $\tilde{D}(x)$ has dropped to 
\begin{equation}
\label{eq:Dscar}
   D_{\text{scar}} =\left( 1 +\lambda \right) D_0,
\end{equation}
where $D_0$ is the value of $\tilde{D}(x)$ in the normal (healthy) region, and 
$-1 < \lambda < 0$. 
For example, for a cable length of $L=3.0$ cm with $D_0=0.005\;$cm$^2$ms$^{-1}$, 
a defected region of length $L_{\text{scar}}=0.5$ cm and $\lambda=-0.8$ would have
$D_{\text{scar}}=0.001$ cm$^2$ms$^{-1}$.
Hence, the spatially dependent voltage diffusion coefficient has the form
\begin{eqnarray}
\label{tilded}
   \tilde{D}(x) = \left \{ 
     \begin{array}{ll}
        D_{scar}, & \mbox{if $x_{scar} < x <  x_{scar} +L_{scar}$} \\
        D_{0},    & \mbox{anywhere else.}
     \end{array}
\right. 
\end{eqnarray} 
Note that in a recent work \cite{Bragard2021}, a spatially and temporally 
diffusion coefficient was considered which encompasses conductance 
heterogeneities in the cardiac tissue induced by the dynamics of the gap 
junctions. Obviously, the adjustable parameters of the voltage diffusion profile
is the starting point of the defected region $x_{\text{scar}}$, the spatial 
length of the scar tissue $L_{\text{scar}}$, and the percentage decrease
$-\lambda$ which lowers $\tilde{D}(x)$ in the defected region. Using 
Eqs. (\ref{eq:Dscar}) and (\ref{tilded}) above, the spatially averaged diffusion 
coefficient is
\begin{equation}
\label{tildedave}
   <\tilde{D}(x)> =D_0 \left( 1 +\lambda \frac{L_{\text{scar}}}{L} \right).
\end{equation}

%%%%%%%%%%%%%%%%%%%%%%%%%%%%%%%%%%%%%%%%%%%%%%%%%%%%%%%%%%%%%%%%%%%%%%%%%%%%%%%%
\subsection{Numerical Calculations}
All numerical simulations for Eqs.\;(\ref{eq:eq01}) - (\ref{eq:eq03}) along 
with Eqs.\;(\ref{eq:eq06}) - (\ref{eq:eq08}) were performed on a theoretical 
1D cable of cells, using the fourth order Runge-Kutta algorithm with fixed 
time-step $dt = 0.002 \text{\;ms}$. For spatially discretizing 
Eqs. (\ref{eq:eq01}) - (\ref{eq:eq03}), the spatial domain was divided into 
$N_x -1$ elements with $N_x =400$ nodes at $x_i =(i-1) L / (N_x -1)$ 
($i=1,2, ...,N_x$) which are separated by distance 
$dx = L/(N_x -1) =0.0075$ cm (about the length of a cardiac cell). 
Second order, centered finite difference formulas were used to discretize the 
first and second derivatives of the state variables wherever they appear in 
Eqs. (\ref{eq:eq01}) - (\ref{eq:eq03}). The spatially discretized equations 
are given explicitly in the next subsection. For numerical purposes, the 
spatially dependent (inhomogeneous) diffusion coefficient $\tilde{D}(x)$ is
modeled as a double step-function controlled by two very steep $\texttt{tanh}$ 
functions. Independent runs were executed using different numerical codes 
written in Python and Fortran 95, and the results were verified to be 
practically the same. Unless otherwise specified, the number of time-steps 
were $150,000$.

The boundary conditions at the ends of the cable are chosen to be those of 
zero-flux (Neumann) type, i.e., 
\begin{equation}
\label{eq:eq12}
   \tilde{D}(x) \left. \frac{\partial u(x,t)}{\partial x} \right|_{x=0}
  =\tilde{D}(x) \left. \frac{\partial u(x,t)}{\partial x} \right|_{x=L}
  =0,
\end{equation}
where $L$ is the length of the cable which, in what follows, is set everywhere 
equal to $3$ cm ($L=3.0$ cm). As explained above, the (inhomogeneous) diffusion
coefficient $\tilde{D}(x)$ is practically a piece-wise constant function which 
assumes the value $D_0$ and $D_{\text{scar}}$ in the normal (healthy) and 
defected (scar) tissue region, respectively, as it is shown schematically in 
Fig. \ref{fig:Dtilde}.

%%%%%%%%%%%%%%%%%%%%%%%%%%%%%%%%%%%%%%%%%%%%%%%%%%%%%%%%%%%%%%%%%%%%%%%%%%%%%%%%
For a homogeneous (spatially constant) diffusion coefficient $\tilde{D}(x) =\tilde{D}$
along the cable, Eq. (\ref{eq:eq01}) becomes
\begin{equation}
\label{eq10}
\frac{\partial u}{\partial t} = \tilde{D} \cdot \frac{\partial^2 u}{\partial x^2}  
                 - J_{\text{fi}} - J_{\text{so}} - I_{\text{si}} + J_{\text{stim}} \;.
\end{equation}
If the diffusion coefficient $\tilde{D}$ is allowed to depend on the spatial
coordinate $x$, the first term on the right hand side of Eq. (\ref{eq:eq01}), 
i.e., $\nabla \left( \tilde{D} \nabla u \right)$ becomes 
\begin{equation}
\label{eq:eq11}
   \frac{\partial}{\partial x} 
    \left[ \tilde{D}(x) \frac{\partial u(x)}{\partial x} \right]
  =\frac{\partial \tilde{D}(x)}{\partial x} \frac{\partial u(x,t)}{\partial x} 
   +\tilde{D}(x) \frac{\partial^2 u(x,t)}{\partial x^2}\;.
\end{equation} 
From  Eq. (\ref{eq:eq11}) we can see that we need the discrete form of the 
first and second spatial derivative of $u(x,t)$, as well as the first spatial 
derivative of $\tilde{D}(x)$. We use the following centered differences
\cite{LangtangenHansPetter2017FDCW}
\begin{align}
\label{eq:eq13}
   \frac{\partial u(x,t)}{\partial x} &=\frac{u_{i+1}(t) -u_{i-1}(t)}{2\Delta x},
\\
\label{eq:eq14}
   \frac{\tilde{D}(x)}{\partial x} &=\frac{\tilde{D}_{i+1} -\tilde{D}_{i-1}}{2\Delta x},
\\
\label{eq:eq15}
   \frac{\partial^2 u(x,t)}{\partial x^2} 
   &=\frac{u_{i+1}(t) -2 u_{i}(t) +u_{i-1}(t)}{\Delta x^2}.
\end{align}
Using Eqs.\;(\ref{eq:eq11}-\ref{eq:eq13}), the spatially discretized system
of Eqs.\;(\ref{eq:eq01}-\ref{eq:eq03}) reads
\begin{align}
\label{eq:eq16}
   \frac{\partial u_i}{\partial t} &=
    \frac{\tilde{D}_{i+1} -\tilde{D}_{i-1}}{2\Delta x}
     \frac{u_{i+1} -u_{i-1}}{2\Delta x}
     +\tilde{D}(x_i) \frac{u_{i+1} -2 u_{i} +u_{i-1}}{\Delta x^2}
\nonumber \\
    &-J_{\text{fi}}(u_i;v_i) -J_{\text{so}}(u_i) -J_{\text{si}}(u_i;w_i) +J_{\text{stim}}(x_i,t),\;\;
\\
\label{eq:eq17}
   \frac{\partial v_i}{\partial t} &=\Theta(u_c-u_i) \frac{1-v_i}{\tau_v^-(u_i)}
                                    -\Theta(u_i-u_c) \frac{v_i}{\tau_v^+},\;\;
\\
\label{eq:eq18}
   \frac{\partial w_i}{\partial t} &=\Theta(u_c-u_i) \frac{1-w_i}{\tau_w^-}
                                   -\Theta(u_i-u_c) \frac{w_i}{\tau_w^+},\;\;
\end{align}
where it is implied that the discretized variables $u_i$, $v_i$, and $w_i$ 
depend on time $t$. The stimulus current, which is necessary for the 
excitation of the AP pulse, is assumed to arise from physiological 
mechanisms of the heart.
There is a large volume of works on the calculation of the ventricular AP in 
1D \cite{Lesh1989,Cain2004,Oliver2005,Penaranda2012} using various types of 
stimulus current functions $J_{\text{stim}}(x,t)$. The FK3V model has been 
also used for the calculation of the ventricular AP in two and three dimensions 
\cite{Fenton2002}. Also, mapping models have been used for the analysis of 
numerical results obtained through the FK3V model \cite{Tolkacheva2002}.

%%%%%%%%%%%%%%%%%%%%%%%%%%%%%%%%%%%%%%%%%%%%%%%%%%%%%%%%%%%%%%%%%%%%%%%%%%%%%%%%
\begin{figure}[!t]
\centering
   \includegraphics[scale=0.25]{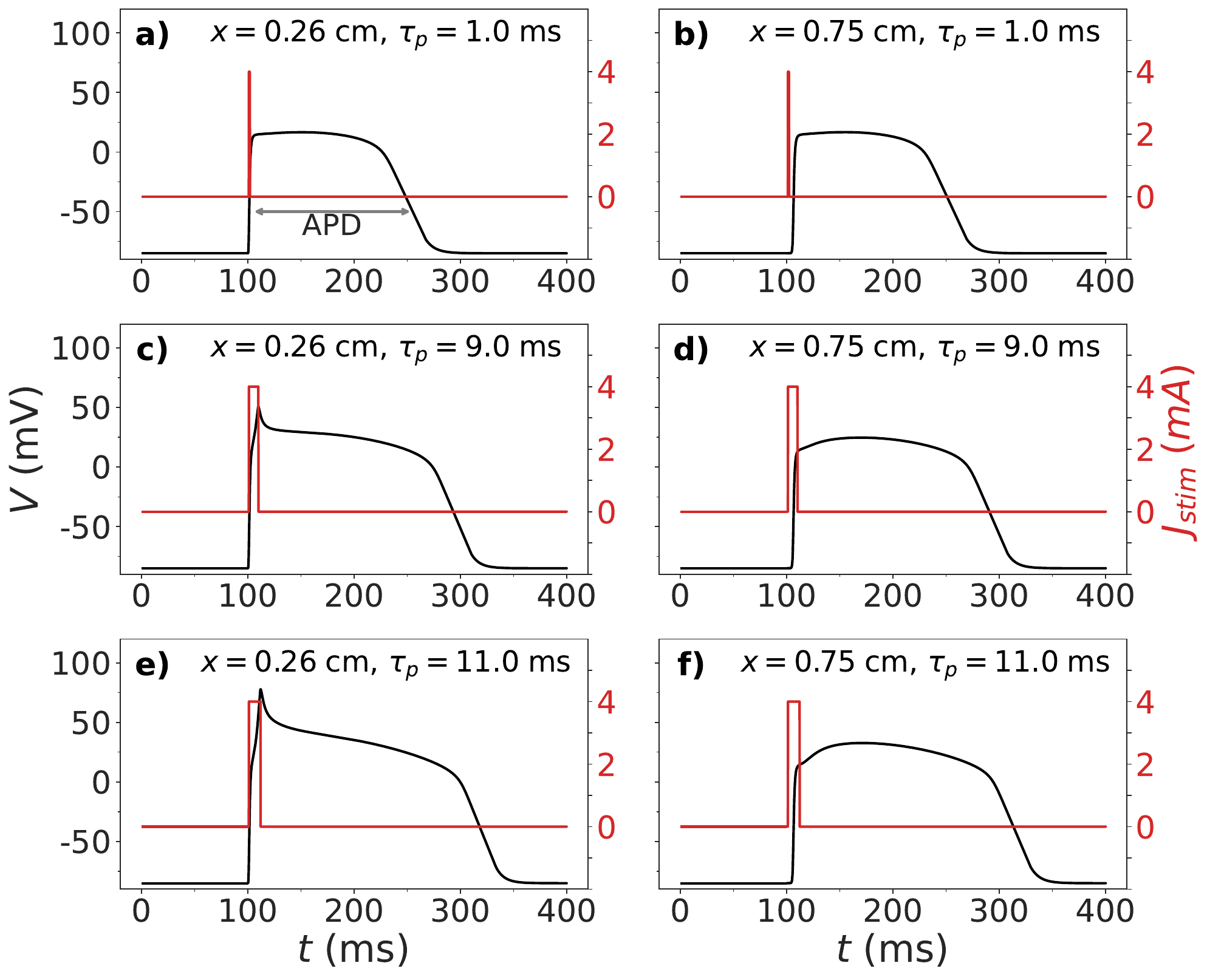}
   \caption{The action potential $V$ (black curves) in natural units as a function 
   of time $t$ that is produced by a single rectangular current pulse depicted by 
   the red curve (notice the small deflection in the beginning) for 
   $D=0.005\;$cm$^2$ms$^{-1}$, $L_{\text{exc}}=0.113$ cm, $J_{\text{{amp}}}=4.0$ mA. 
   The action potential duration (APD), indicated by the gray horizontal 
   double-headed arrow, is measured for each plot. 
   (a,b) $\tau_{\text{p}}=1.0\;$ms, and APD$=159\;$ms; 
   (c,d) $\tau_{\text{p}}=9.0\;$ms, and APD$=194.4\;$ms; 
   (e,f) $\tau_{\text{p}}=11.0\;$ms, and APD$=215.8\;$ms;.  
   The action potential is monitored at positions $x\simeq 0.26\;$cm (right panels), 
   and at $x\simeq 0.75\;$cm (left panels).
}
\label{fig:tps}
\end{figure}
%%%%%%%%%%%%%%%%%%%%%%%%%%%%%%%%%%%%%%%%%%%%%%%%%%%%%%%%%%%%%%%%%%%%%%%%%%%%%%%%

%%%%%%%%%%%%%%%%%%%%%%%%%%%%%%%%%%%%%%%%%%%%%%%%%%%%%%%%%%%%%%%%%%%%%%%%%%%%%%%%
\begin{figure}[!t]
   \includegraphics[scale=0.26]{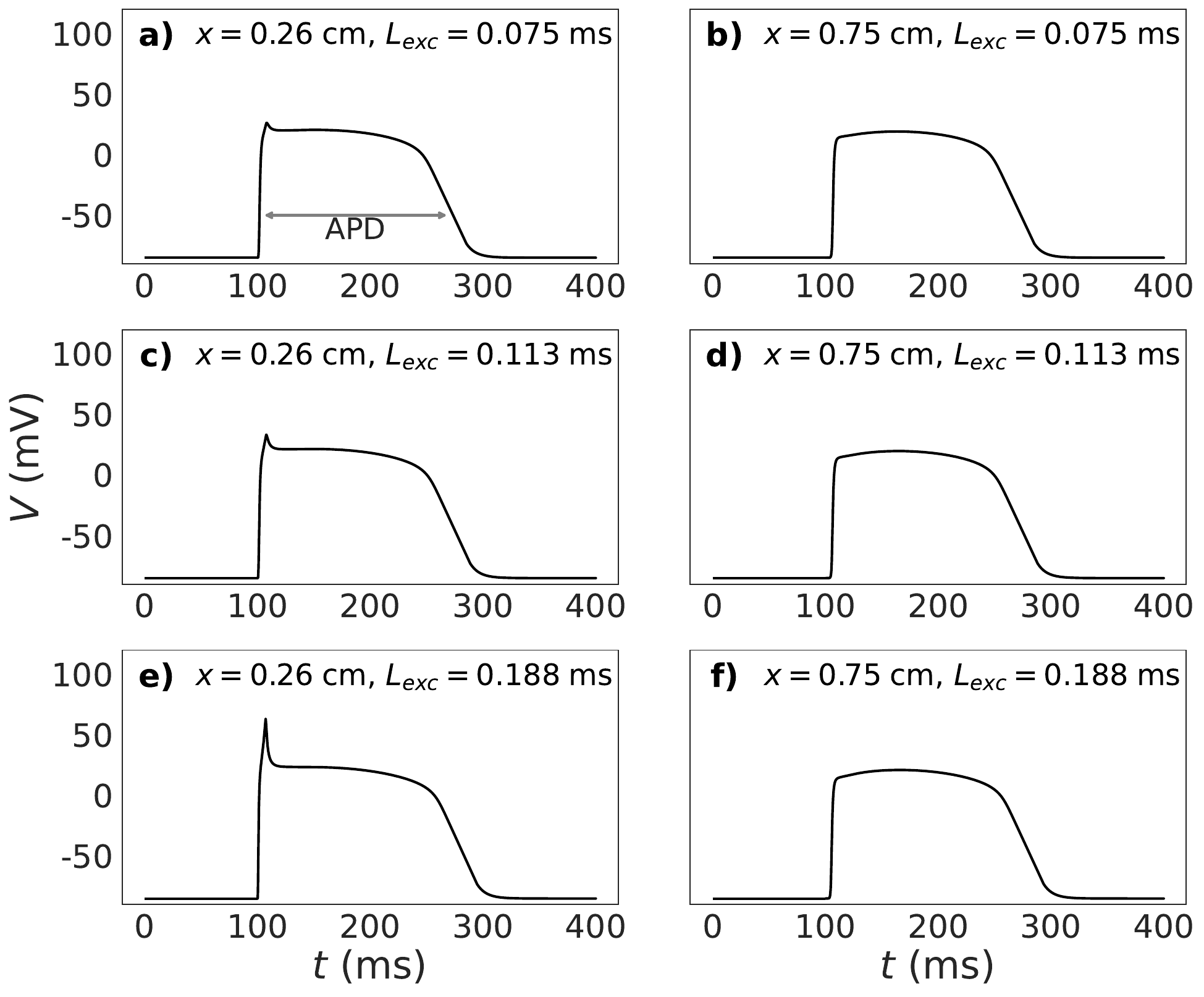}
   \caption{The action potential $V$ (black curves) in natural units as a function 
            of time $t$ excited by a single rectangular pulse current with amplitude 
            $J_{\text{amp}}=4\;$mA, and duration $\tau_{\text{p}} =7\;$ms (not shown). 
            The action potential duration (APD), indicated by the gray horizontal 
            double-headed arrow, is measured for each plot.
           (a,b) $L_{\text{exc}}=0.075~$cm, and APD$=173.9$ms; 
           (c,d) $L_{\text{exc}}=0.113~$cm, and APD$=174.9$ms; 
           (e,f) $L_{\text{exc}}=0.188~$cm, and APD$=179.5$ms.
           The action potential $V$ is monitored at $x\simeq 0.25~$cm 
           (right panels), and at $x\simeq 0.75~$cm (left panels).
}
\label{fig:fig03}
\end{figure}
%%%%%%%%%%%%%%%%%%%%%%%%%%%%%%%%%%%%%%%%%%%%%%%%%%%%%%%%%%%%%%%%%%%%%%%%%%%%%%%%

%%%%%%%%%%%%%%%%%%%%%%%%%%%%%%%%%%%%%%%%%%%%%%%%%%%%%%%%%%%%%%%%%%%%%%%%%%%%%%%%
\section{Results}
\subsection{The Action Potential}
Using Eq.\;(\ref{eq:eq16}-\ref{eq:eq18}), we have calculated numerically the 
ventricular AP propagating through ventricular tissue of length $L=3 $cm as a 
function of time $t$. A small segment of the tissue/cable of length 
$L_{\text{exc}}=0.11$ cm is initially excited through its left end, i.e., the 
segment from $x=0$ to $x =L_{\text{exc}}=0.11~cm$, using stimulus currents of 
amplitude $J_{\text{amp}}=5$ mA and different durations $\tau_{\text{p}}$. 
Typical AP pulse profiles (black curves) along with the associated stimulus 
currents (red curves) are shown in Fig. \ref{fig:tps}, monitored at two 
different locations on the cable, i.e., at $x\simeq 0.26~cm$ (relatively close to 
the excited region, left panels) and $x\simeq 0.75~cm$ (at one-fourth of the 
cable length as measured from $x=0$, right panels). As it can be observed, the 
amplitude of the AP as well as its duration (action potential duration, APD) 
increases with increasing $\tau_p$ (from top to bottom). The latter, specifically, 
which is defined as the width of the pulse at $12\%$ of its maximum amplitude 
(illustrated in (a) by the gray horizontal double-headed arrow), increases from 
$159$ ms for $\tau_{\text{p}}=1.0$ ms, to $194.4$ ms for $\tau_{\text{p}}=9.0$ ms, 
to $215.8$ ms for $\tau_{\text{p}}=9.0$ ms. Left and right panels, obtained by 
monitoring the AP pulses at different locations on the cable, also differ in 
that the former exhibit a sharp peak at a time instant corresponding to the end 
of the stimulus current pulse. This sharp peak decreases until it practically 
vanishes for locations on the cable relatively far from the excited region.

Similarly, in Fig. \ref{fig:fig03}, the calculated action potential $V$ as 
a function of time $t$ is monitored at two different positions on the cable for 
stimulus currents of amplitude $J_{amp} =4~mA$, duration $\tau_p =4~ms$, and 
three different values of the initially excited segment at the left end of the 
cable of length $L_{\text{exc}}$, which extends from $x=0$ to $x=L_{\text{exc}}$.
As in Fig. \ref{fig:tps}, the action potential is monitored at $x\simeq 0.25~cm$ 
and $x\simeq 0.75~cm$ (left and right panels, respectively). 
Again it is observed that, the duration of the action potential (APD) increases 
with increasing $L_{\text{exc}}$. Specifically, the APD increases from $173.9$ ms
for $L_{\text{exc}}=0.075$ cm to $174.9$ ms for $L_{\text{exc}}=0.113$ cm, to 
$179.5$ for $L_{\text{exc}}=0.113$ cm.
In both Figs. \ref{fig:tps} and \ref{fig:fig03}, the action potential exhibits 
the right characteristics in (e) and (f) panels, as long as the shape and the 
width (i.e., the APD) is concerned.

%%%%%%%%%%%%%%%%%%%%%%%%%%%%%%%%%%%%%%%%%%%%%%%%%%%%%%%%%%%%%%%%%%%%%%%%%%%%%%%%
\subsection{The pseudo-ECG}
The analysis and interpretation of ECGs remains mostly empirical. The pseudo-ECG 
at a particular time-instant $t$ is calculated numerically from the spatial profile
of the AP at that time-instant on the cable using the expression 
\cite{Gima2002, Clayton2004, Aslanidi2005,  Wang2006, Bueno2008} 
(for a thorough derivation see Ref. \cite{PlonseyRobert2007BAQA})
\begin{equation}
\label{eq:eq19}
   \Phi_e ({\bf x}^\star, t) 
   =-K \int \nabla V({\bf x},t) \cdot \nabla \frac{1}{|{\bf x}^\star -{\bf x}|} d{\bf x}, 
\end{equation}
where $\nabla V({\bf x},t)$ is the spatial gradient of the ventricular AP, 
$K=1.89\text{\;mm}^2$ is a constant that depends on electrophysiological quantities, 
such as the radius of the fiber and the intracellular conductivity. The ``electrode'' 
measuring the voltage is at point ${\bf x}^\star$ of the fiber, and 
$|{\bf x}^\star -{\bf x}|$ is the distance from a source point $\bf x$ to a field 
point ${\bf x}^\star$ (${\bf x}^\star > {\bf x}$. The temporal profile of the 
pseudo-ECG $\Phi_e$ constitutes an approximation for the ventricular component of 
the ECG, i.e., the pseudo-ECG generated at a hypothetical electrode which is 
located at a particular distance away from the last epicardial cell along the cable. 
As shown in Fig.\;(\ref{fig:Conduction}), the ventricular potential contributes 
specifically to the formation of the QRS cluster and the T wave. The pseudo-ECG is, 
thus, expected to reproduce these features.

\begin{figure}[!t]
\includegraphics[scale=0.6]{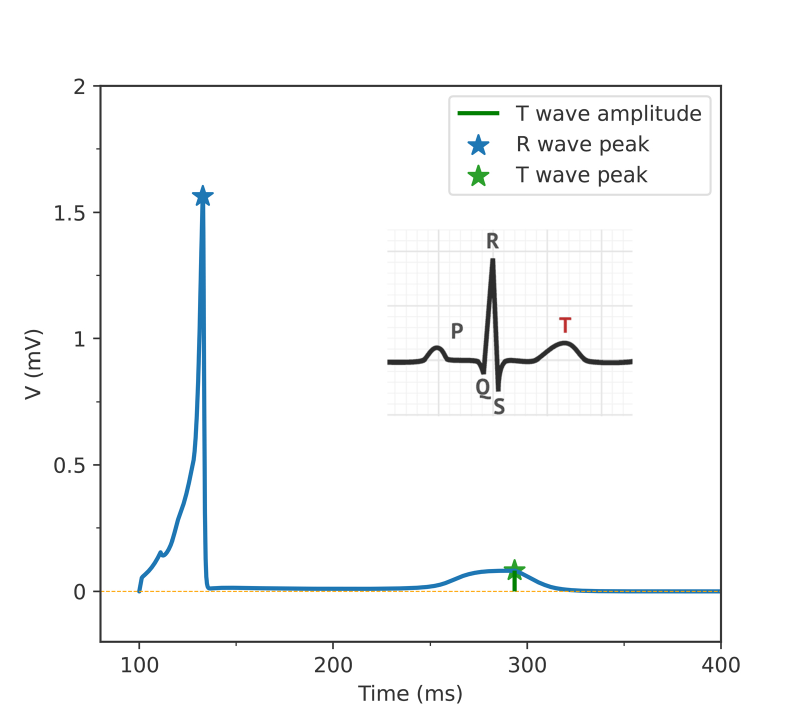}
\caption{Simulated pseudo-ECG as a function of time $t$ calculated using the 
three-variable Fenton-Karma model . For comparison, a drawing of a real ECG is 
shown in the inset. We can clearly detect the R and T wave equivalents whose 
amplitudes we designate with the blue and green stars respectively.
}
\label{fig:pseudo}
\end{figure}

In one dimension, Eq. (\ref{eq:eq19}) reads
\begin{equation}
\label{eq20}
   \Phi_e (x^\star,t) =-K \int \frac{\partial V(x,t)}{\partial x} 
    \left( \frac{\partial}{\partial x} \frac{1}{|x^\star -x|} \right) dx~.
\end{equation}
We calculate the pseudo-ECG for point outside the cell cable, so for $x^\star>L$, 
Eq. (\ref{eq20}) becomes
\begin{equation}
\label{Phi}
   \Phi_e (x^\star,t) =-K \int \frac{\partial V(x,t)}{\partial x} 
     \; \frac{1}{(x^\star -x)^2} \;dx,
\end{equation}
and is thus more easily calculated. Using the spatial profiles calculated from 
Eq.\;(\ref{eq:eq16}) - (\ref{eq:eq18}) at each time instant, we calculate 
$\Phi_e (L,t)$ which is the desired pseudo-ECG; in our calculations 
$x^\star=3.37$ cm, while the cell cable length is $L=3.0$ cm. 
As shown in Fig. \ref{fig:pseudo}, the T-wave has positive polarity and its 
amplitude is defined as the vertical distance from $V=0\;$. In general, T waves 
are considered positive when their deflection is upward, and negative when it
is downward. For {\em biphasic T-waves} (waves with both an upward and a downward 
deflection), unless otherwise stated, the dominant deflection is chosen. In the 
small inset, a surface ECG which is recorded using two electrodes placed on the 
skin surface, away from the heart, is visually compared to the pseudo-ECG.

%%%%%%%%%%%%%%%%%%%%%%%%%%%%%%%%%%%%%%%%%%%%%%%%%%%%%%%%%%%%%%%%%%%%%%%%%%%%%%%%
%%%%%%%%%%%%%%%%%%%%%%%%%%%%%%%%%%%%%%%%%%%%%%%%%%%%%%%%%%%%%%%%%%%%%%%%%%%%%%%%

%%%%%%%%%%%%%%%%%%%%%%%%%%%%%%%%%%%%%%%%%%%%%%%%%%%%%%%%%%%%%%%%%%%%%%%%%%%%%%%%
\begin{figure}[!t]
\centering
\includegraphics[scale=0.65]{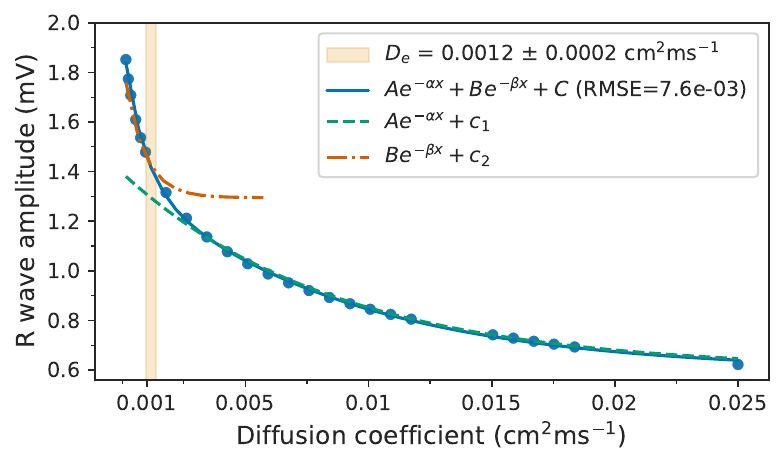}
\caption{R-wave amplitude as a function of the (homogeneous, spatially constant) 
         voltage diffusion coefficient $\tilde{D} =D_0$, extracted from the 
         calculated pseudo-ECGs using Eq. (\ref{eq20}) for a total of $25$ values 
         for $D_0$ within the interval $0.0005 - 0.025~\rm{cm}^2\rm{ms}^{-1}$. 
         Other simulation parameters are: 
         $J_{\text{amp}}=10.0$ mA, $t_{\text{p}}=10.0$ ms, 
         $L_{\text{exc}}=0.105$ cm, dx = 0.0075 cm, dt = 0.0013 ms, 
         Nsteps = 230769. 
         Curve fit parameters are: 
         A = 0.79, $\alpha$ = 115.36, B = 0.56, $\beta$ = 1258.84, and C = 0.595. 
         The range of values $D_{\text{e}} =$ 
         (0.0012 $\pm$ 0.0002)\;cm$^2$ms$^{-1}$ denoted by the shaded area, is a 
         range of experimental values used frequently in the literature. 
         The constants for the failed single exponentials are $\text{c}_1=0.6$ 
         (green curve), and $\text{c}_2=1.3$ (red curve).}
\label{fig:DvsRexp}
\end{figure} 
%%%%%%%%%%%%%%%%%%%%%%%%%%%%%%%%%%%%%%%%%%%%%%%%%%%%%%%%%%%%%%%%%%%%%%%%%%%%%%%%

%%%%%%%%%%%%%%%%%%%%%%%%%%%%%%%%%%%%%%%%%%%%%%%%%%%%%%%%%%%%%%%%%%%%%%%%%%%%%%%%
\begin{figure*}[!t]
   \includegraphics[scale=0.42]{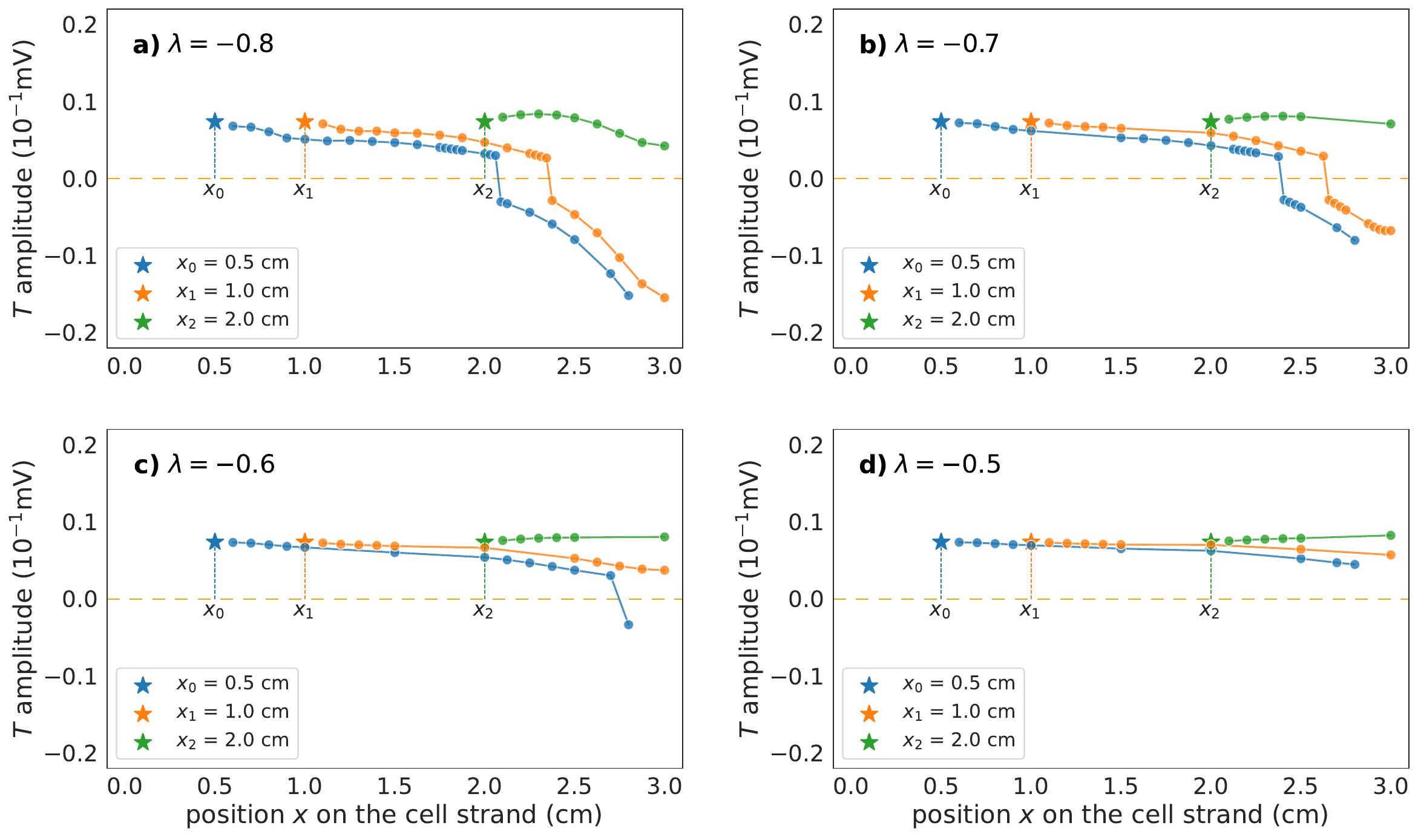}
   \caption{T-wave morphology depicted as magnitude of the peak or the dip of 
   the T-wave (in units of mV) for various starting 
   positions and widths of the defected tissue. The variable $x$ ($0<x<L$) is the 
   position on the cell strand. The star at each $x_j$ ($j=0,1,2$) marks the 
   position of the beginning of the defect for three different values of 
   $x_{\text{scar}} =0.5$, $1.0$, and $2.0$. All results are color-coded by 
   this value. The location of each dot on the $x-$axis signifies 
   the end point of the defected region at $x_{\text{scar}} +L_{\text{scar}}$, 
   i.e., the distance along the $x-$axis of the dot from the star on the same 
   curve represents the value of $L_{\text{scar}}$. Each plot has a different 
   value for the parameter $\lambda$ as indicated in the label on top. For all 
   four subplots, the effective diffusion coefficient is 
   $D_0=0.005\;$cm$^2$ms$^{-1}$.}
\label{fig:Da_horiz}
\end{figure*}
%%%%%%%%%%%%%%%%%%%%%%%%%%%%%%%%%%%%%%%%%%%%%%%%%%%%%%%%%%%%%%%%%%%%%%%%%%%%%%%%

\subsection{Constant Diffusion Coefficient}

\indent We first run our model with a spatially constant diffusion coefficient 
$\tilde{D} =D_0$. This can be regarded as an effective parameter, a mean value 
to account for the discontinuity defect part inserts. The height of the R-wave 
(see blue star in Fig. \ref{fig:pseudo}) in each pseudo-ECG denotes the value 
of the T-wave amplitude in units of mV; when plotted for $25$ different values, 
as shown in Fig. \ref{fig:DvsRexp}, it appears to exhibit an exponential 
dependence on the effective diffusion coefficient $\tilde{D}$. 
Since repeated attempts to fit a single exponential curve using least squares 
failed, we tried using the sum of two exponentials via the {\em ansatz}
\begin{equation}
\label{ansatz}
   R_{wa} =A\;e^{-\alpha \tilde{D} } + B\;e^{-\beta \tilde{D}} +\;C, 
\end{equation}
where $R_{wa}$ is the R-wave amplitude, and $A$, $\alpha$, $B$, $\beta$, and $C$, 
are parameters to be fitted. Using the ansatz (\ref{ansatz}), we obtained 
excellent fit using parameters $A = 0.79$, $\alpha =115.36$, $B =0.56$, 
$\beta =1258.84$, and $C =0.595$. In the exemplary fit shown by the blue line in 
Fig. \ref{fig:DvsRexp}, we notice a transition region around the value of 0.0012 
$\pm$ 0.0002\;cm$^2$ms$^{-1}$ of the diffusion coefficient, which, as mentioned 
before, is an experimental value used frequently in literature. 
The transition region is identified by those values of $\tilde{D} =D_0$ for which 
the fitted single-exponential curves $A\;e^{-\alpha \tilde{D} } +\;C_1$ 
(green-dashed curve) and $B\;e^{-\beta \tilde{D}} +\;C_2$ (red dashed-dotted curve) 
start diverging significantly from the numerical data (slightly above the experimental 
value of $\tilde{D} =D_e$. 
The algorithm was implemented using the \texttt{SciPy} python library.

%%%%%%%%%%%%%%%%%%%%%%%%%%%%%%%%%%%%%%%%%%%%%%%%%%%%%%%%%%%%%%%%%%%%%%%%%%%%%%%%
\subsection{Modeling a Space-Dependent Diffusion Coefficient}

To study the polarization of the T-wave in a tissue containing a localized 
defect, a stimulus current in the form of a rectangular pulse of amplitude 
$J_{\text{amp}} =0.9$ mA, and duration $\tau_{\text{p}}=11$ ms, was applied at 
the first $15$ cells of the cable, those which are at its left end ($x=0$), 
whose length is $L_{\text{exc}} \simeq 0.011$ cm. Then, the pseudo-ECG is 
calculated from the spatio-temporal profile of the APs, and the maximum magnitude
of the T-wave is identified. This procedure was repeated as a function of the width
of the defect $L_{\text{scar}}$ for three different values of the position of the 
onset of the defect $x_{\text{scar}}$ and four values of the parameter $\lambda$.
The results are presented in a compact way in Fig.\;\ref{fig:Da_horiz}. In all four
subfigures, the diffusion coefficient in the healthy region is 
$D_0 =0.005$ cm$^2$ ms$^{-1}$.
The defect was modeled using a spatially dependent diffusion coefficient $\tilde{D}(x)$, 
whose characteristics were previously depicted in Fig. \ref{fig:Dtilde}. 
For the results presented in Fig. \ref{fig:Da_horiz}, the defected region spans 
the interval from $x =x_{\text{scar}}$ to $x =x_{\text{scar}} +L_{\text{scar}}$. 
Within this interval, the diffusion coefficient is $\tilde{D} =L_{\text{scar}}$,
with $\lambda=-0.8$, $-0.7$, $-0.6$, and $-0.5$ in Fig.\;\ref{fig:Da_horiz}(a), (b),
(c), and (d), respectively. Obviously, the relation 
$x_{\text{scar}} +L_{\text{scar}} < L$ should hold in any case.

By inspection of Fig.\;\ref{fig:Da_horiz} we observe that the curves for 
$x_{\text{scar}} =x_3 =2$ cm (green curves) always remain on the positive side 
of the vertical axis, meaning that in this case there is no polarization
inversion of the corresponding T-wave, and thus this is always positive.
The same holds true for any other value of $x_{\text{scar}} > 2$, since the
defected region constitutes only a relatively small part of the cable, which
is of length $L =3$ cm, so that it cannot affect significantly the spatio-temporal
AP profile.
It can be also be observed from Fig.\;\ref{fig:Da_horiz}(d) that all three curves
remain on the positive sides of the vertical axis, and thus no T-wave inversion 
appears, due to the relatively small magnitude of $\lambda$. Indeed, the magnitude 
of $\lambda$ in this case does not seem to be sufficiently high (or equivalently 
the defect is not sufficiently deep) to invert T waves. 
For slightly deeper defect, for $\lambda =-0.6$, as shown in 
Fig.\;\ref{fig:Da_horiz}(c), T-wave inversion is observed for 
$x_{\text{scar}} =x_0 =0.5$ and $L_{\text{scar}} =2.25$ (blue curve) but not 
for $x_{\text{scar}} =x_1 =1.0$ or $x_{\text{scar}} =x_2 =2.0$ (orange and green
curves, respectively). The obvious reason is that in the latter cases the 
width of the defected region $L_{\text{scar}}$ cannot reach such a high value as
that in the former case ($L_{\text{scar}} =2.25$). Moreover, as it can be observed
from Figs.\;\ref{fig:Da_horiz}(a) and (b) the parts of the curves with 
$x_{\text{scar}} =x_0 =0.5$ (blue curves) and $x_{\text{scar}} =x_1 =1.0$ 
(orange curves), respectively, with inverted (negative) T-wave become larger 
with decreasing $\lambda$. From these observations can thus be concluded that 
for fixed $x_{\text{scar}}$, deep and wide defected regions favor T-wave inversion. 
Furthermore, the value of $L_{\text{scar}}$ at the transition from positive to negative 
T waves is lower in the orange curves ($x_{\text{scar}} =x_1 =1.0$) than that in the 
blue curves ($x_{\text{scar}} =x_0 =0.5$) as can be observed from 
Figs.\;\ref{fig:Da_horiz}(a) and (b). Thus, for fixed $\lambda$, defects with 
higher $x_{\text{scar}}$ are capable to invert T-waves with lower $L_{\text{scar}}$.
From the above remarks it becomes clear that the width, the depth, and the starting
position of the defect contribute decisively to T-wave morphology.

%%%%%%%%%%%%%%%%%%%%%%%%%%%%%%%%%%%%%%%%%%%%%%%%%%%%%%%%%%%%%%%%%%%%%%%%%%%%%%%%
\begin{figure}[h]
\includegraphics[scale=0.4]{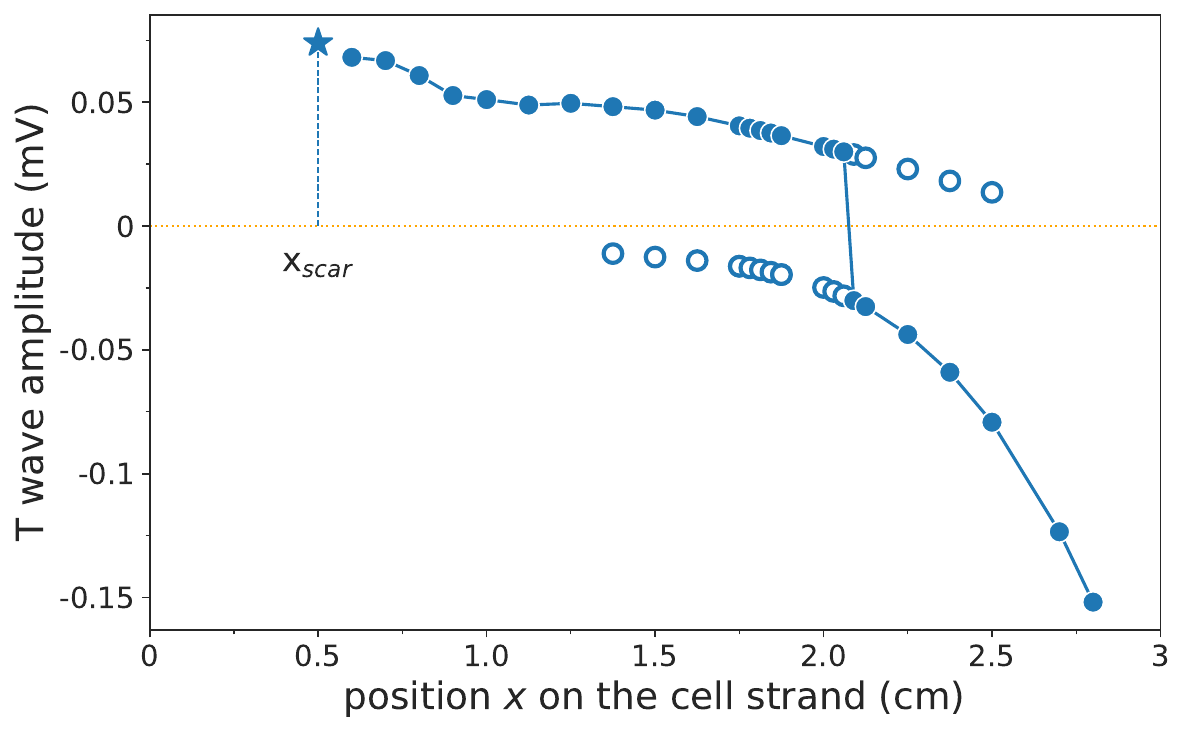}
\caption{T-wave maximae and minimae as a function of the width of the defected 
        region $L_{\text{scar}}$, as obtained from the pseudo-ECG with inhomogeneous 
        voltage diffusion coefficient $\tilde{D}(x)$ with $D_0 =0.005$ cm$^2$ms$^{-1}$
        and $\lambda =-0.8$. The defected regions start at $x_{\text{scar}} =0.5\;$cm.
        The blue solid curve is a guide to the eye. The representation is the same 
        as that in Fig.\;\ref{fig:Da_horiz}. The points depicted as blue solid circles 
        are the ones chosen as the largest of the two in the biphasic wave. 
        The hollow white circles show the amplitude of the other wave in the 
        biphasic phase.  The truly biphasic 
        phase is limited to a few points around the blue vertical segment at 
        $L_{\text{scar}} \simeq 1.6$. The purpose of this plot is to show that 
        there is a transition phase during T inversion, where the wave has both 
        positive (upward) and negative (downward) parts.
}
\label{fig:Da_box}
\end{figure} 
%%%%%%%%%%%%%%%%%%%%%%%%%%%%%%%%%%%%%%%%%%%%%%%%%%%%%%%%%%%%%%%%%%%%%%%%%%%%%%%%

We should note that the transition from positive to negative T waves is realized 
through a biphasic stage, with a minimum and a maximum of similar magnitude. 
This is consistent with the bibliography where it is reported that biphasic T 
waves usually evolve and are often followed by T-wave inversion with strongly 
suspected myocardial ischaemia \cite{ChannerKevin2002Mi}. There was no attempt made 
to trace the biphasic stage in Fig.\;\ref{fig:Da_horiz}, which is actually limited
within a small interval around the transition point. Wherever two extremae appear
in the calculated T-wave, only the higher of them is plotted. However, a typical 
biphasic stage of the calculated T-wave is illustrated below. 

%%%%%%%%%%%%%%%%%%%%%%%%%%%%%%%%%%%%%%%%%%%%%%%%%%%%%%%%%%%%%%%%%%%%%%%%%%%%%%%%
\begin{figure*}[!t]
\includegraphics[scale=1.0]{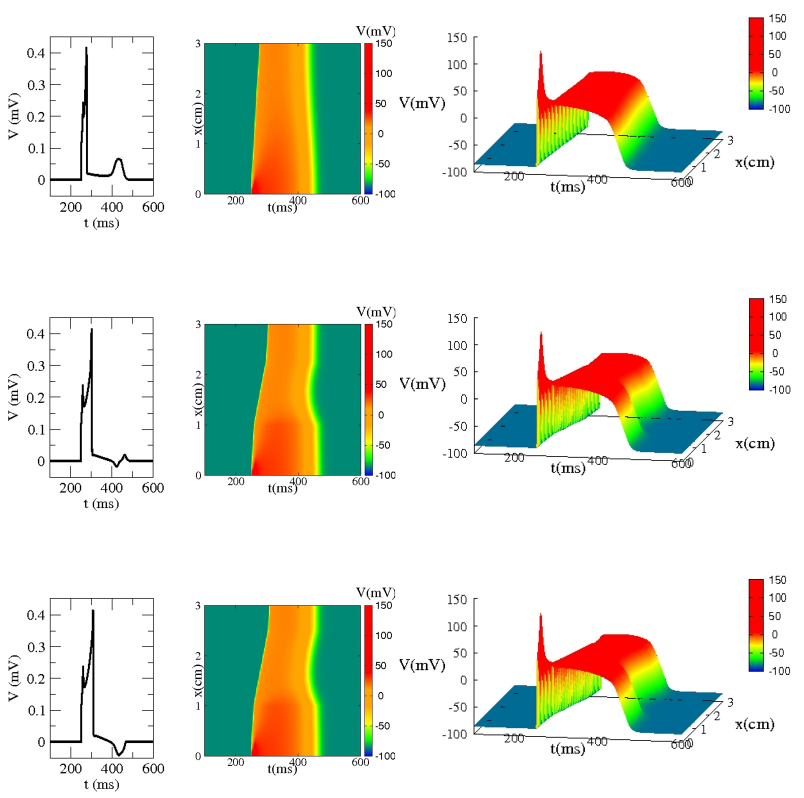}
\caption{Pseudo-ECGs as a function of time $t$, two-dimensional maps of the 
         action potential on the $x - t$ plane, and three dimensional plots of 
         the action potential on the $x - t$ plane, are shown on the left, middle, 
         and right columns, respectively. The parameters, from top to bottom row
         are $L_{\text{scar}} =0$ (first row), $L_{\text{scar}} =1.25$ cm (second
         row), and $L_{\text{scar}} =1.5$ cm (third row).
         The first, second, and third row show the case of pseudo-ECG with 
         positive, biphasic, and negative (inverted) T-wave, respectively. These
         figures illustrate the effect of the defected region on the 
         spatio-temporal profile of the action potential. The defected region in
         the second and third row starts at $L_{\text{scar}} =0.5$, and the 
         length of the cable is $L=3$ cm in all three rows.
}
\label{fig:T-wave}
\end{figure*} 
%%%%%%%%%%%%%%%%%%%%%%%%%%%%%%%%%%%%%%%%%%%%%%%%%%%%%%%%%%%%%%%%%%%%%%%%%%%%%%%%

In Fig.\;\ref{fig:Da_box}, the maximum and the minimum of the T-wave (equivalently 
the maximum and the second maximum of the magnitude of the T-wave) are plotted 
as a function of the width of the defected region $L_{\text{scar}}$, for the 
parameters of the 
blue curve in Fig.\;\ref{fig:Da_horiz}(a). Recall that all points on that curve 
were obtained from the pseudo-ECG using voltage diffusion coefficient $\tilde{D}$
with $D_0 =0.005$ cm$^2$ms$^{-1}$, $\lambda =-0.8$, and a defected region starting 
at $x_{\text{scar}} =x_0 =0.5$ cm. The blue circles (filled and empty) have been
obtained through numerical calculations while the (blue) solid curve is a guide 
to the eye, actually indicating the transition from positive to negative (inverted)
T-wave. The filled and empty circles indicate maximae and second maximae (whenever 
they exist) of the T-wave magnitude. Note that the truly biphasic stage, for 
which the minimum and the maximum of the T-wave have approximately equal magnitude,
is limited to a few ($\sim 5$) points around the blue vertical segment indicating 
the T-wave inversion transition. Further away from that segment, e.g., at 
$x=1.5$ cm, i.e., at $L_{\text{scar}} =1.0$,
the maximum of the T-wave has much larger magnitude of the minimum, and thus the
positive character of the T-wave is dominant. Such cases are regarded as positive
T-waves in Fig.\;\ref{fig:Da_horiz}. Correspondingly, cases in which the negative
(inverted) character of the T-wave is dominant are regarded as inverted T-waves
in Fig.\;\ref{fig:Da_horiz}.
%%%%%%%%%%%%%%%%%%%%%%%%%%%%%%%%%%%%%%%%%%%%%%%%%%%%%%%%%%%%%%%%%%%%%%%%%%%%%%%%

In Fig. \ref{fig:T-wave}, the pseudo-ECG as a function of time $t$, the map of
the action potential on the $x-t$ plane, and the three-dimensional plot of the
action potential on the $x-t$ plane are shown in three different widths 
$L_{\text{scar}}$ of the defected region of the cardiac tissue, to illustrate 
its effect on the spatio-temporal profile of the action potential and eventually 
on the T-wave morphology. In the figure, from the first to third row (from top 
to bottom), the T-wave of the pseudo-ECG is positive, biphasic, and negative 
(inverted), respectively. The results shown on the first row have been obtained 
for a healthy tissue, that is, for an averaged diffusion coefficient $\tilde{D}$ 
which is homogeneous (without defected region, $\lambda=0$). In this case, as 
shown in the map on the second column, the width of the action potential decreases 
monotonically as it propagates from the excitation region of the cable outwards. 
This is also apparent from the three-dimensional plot, where it is also clear that 
the amplitude of the pulse is not significantly affected during propagation. 
The sharp peak of the action potential profile, appearing in all three sub-figures 
in Fig. \ref{fig:T-wave}, is due to the action potential pulse being very close to 
or inside the excitation region of length $L_{exc} =0.11$ cm of the cable. That 
peak however disappears after short time of propagation in all three cases.

The results shown in the second and third row have been obtained with a diffusion 
coefficient $\tilde{D} =\tilde{D}(x)$, as in Eq. (\ref{tilded}), with 
$L_\text{scar} =1.25$ cm and $L_\text{scar} =1.5$ cm, respectively. AS it can be 
observed, the results in the second and third row are significantly affected by
the existence of the defected region. In the second row the T-wave of the 
pseudo-ECG becomes biphasic, while the width of the action potential pulse is not 
any more monotonically decreasing during outward propagation. Instead, the pulse 
narrows substantially and abruptly while it propagates into the defected region,
and becomes wider after departing from it. That effect is also visible in the 
three-dimensional plot, where we may also observe that the amplitude of the 
AP is not significantly affected during propagation, even in the defected region. 
In the third row, the T-wave of the pseudo-ECG is inverted, becoming negative. 
The profile of the propagating AP pulse is in this case very similar to that shown 
in the second row, i.e., it narrows substantially and abruptly when entering the 
defected region and widens again when departing from it. In this case however the 
pulse narrows within a larger interval because of the larger $L_\text{scar} =1.5$ cm. 
The three-dimensional plot is also very similar to that in the second row.

These figures clearly illustrate the effect of the defected region on the 
propagation of the action potential which in turn affect the pseudo-ECG and is
capable of inverting the T-wave. In Fig. \ref{fig:T-wave}, no attempt was made 
to match observed ECG data. This would require to choose the constant $K$ in 
Eq. (\ref{Phi}) and other parameters appropriately. But this is outside the scope 
of this work which aims at showing qualitatively that spatially inhomogeneous 
voltage diffusion coefficients can account for the inversion of the T-wave, and 
also to account for the variation of the R-wave amplitude against a homogeneous 
(constant) diffusion coefficient $\tilde{D} =D_0$.

%%%%%%%%%%%%%%%%%%%%%%%%%%%%%%%%%%%%%%%%%%%%%%%%%%%%%%%%%%%%%%%%%%%%%%%%%%%%%%%%
\section{Conclusions}
We used the ``simple'' FK3V model to simulate the dynamics of the action potential
propagation in a cable and calculate a pseudo-ECG that reproduces the R wave and 
the T wave of an observed ECG. To the best of our knowledge, pseudo-ECG calculation
using the FK3V model has not been reported before. 
Our results connect the propagation of electrical (action) potentials within the 
cardiac tissue with the morphology of the pseudo-ECG, and by extension with what 
physicians actually observe i.e., the ECG. Specifically, our results reveal the
dependence of the R-wave amplitude as a function of the (homogeneous) voltage
diffusion coefficients and, most importantly they point towards an intimate 
relation between inhomogeneous diffusion coefficients (diffusion coefficients 
with defected regions) and the inversion (of the polarity) of the T-wave. The 
latter is often observed in cases of ischemia in ECG recordings by physicians.

Defected regions in the diffusion coefficient represent ``scars'' in the cardiac
tissue where the electrical connection between cells is broken due to destruction
of the gap junctions, as e.g., occurs in ischemia. As a result, the electrical 
conductance is reduced considerably in these defected region(s), leading 
unavoidably to a severe reduction of the diffusion coefficient there. For 
sufficiently large and deep defected regions in the diffusion coefficient, the 
ability of the cardiac tissue to conduct the action potential is strongly affected, 
and as a result the calculated pseudo-ECG exhibits T-wave inversion. Such findings 
in observed ECGs are often related to ischemia.

It should be mentioned, however, that T-wave inversion could be also obtained
using a spatially constant, averaged diffusion coefficient $<\tilde{D}> =D_0$ for 
sufficiently small values of $D_0$. This could represent the case of many small 
scars distributed almost uniformly along the cardiac tissue (the cable), and it is 
a matter of future work. Our approach to use a single-cell wide strand of cardiac 
cells for action potential propagation in a one-dimensional cable was dictated by 
reasons of simplicity.

While 1D numerical simulations capture essential aspects of the cardiac electrical 
action such as its R and T wave morphology, our approach can be certainly extended 
straightforwardly in more dimensions in the future. The results obtained here 
for the T-wave inversion and its dependence on the diffusion coefficient 
configuration (the inhomogeneity) may ignite research on solving the inverse 
problen, i.e., on how to locate a defected/ischemic region in the cardiac tissue
from observed ECG data.
%%%%%%%%%%%%%%%%%%%%%%%%%%%%%%%%%%%%%%%%%%%%%%%%%%%%%%%%%%%%%%%%%%%%%%%%%%%%%%%%

%\clearpage
\begin{acknowledgments}
Computations in this paper were run in part on the FASRC cluster supported by 
the FAS Division of Science Research Computing Group at Harvard University.
GPT and GDB acknowledge support by the research project co-funded by the
Stavros Niarchos Foundation (SNF) and the Hellenic Foundation for Research and Innovation
(H.F.R.I.) under the 5th Call of “Science and Society” Action – “Always Strive for Excellence –
Theodore Papazoglou” (Project Number: 011496).
Authors NL and IK gratefully acknowledge financial support from Khalifa 
University of Science and Technology, Abu Dhabi, United Arab Emirates, via
the project CIRA-2021-064 (8474000412).
\end{acknowledgments}
%%%%%%%%%%%%%%%%%%%%%%%%%%%%%%%%%%%%%%%%%%%%%%%%%%%%%%%%%%%%%%%%%%%%%%%%%%%%%%%%

%%%%%%%%%%%%%%%%%%%%%%%%%%%%%%%%%%%%%%%%%%%%%%%%%%%%%%%%%%%%%%%%%%%%%%%%%%%%%%%%
\end{document}